\documentclass{mn2e}
\usepackage{psfig}
\usepackage{mnras_cite}
\def\gsim{~\rlap{$>$}{\lower 1.0ex\hbox{$\sim$}}}
\def\lsim{~\rlap{$<$}{\lower 1.0ex\hbox{$\sim$}}}

\def\index{\ensuremath{\delta({\rm V}-{\rm I})}}
\def\excess{\ensuremath{({\rm V}-{\rm I})_{\rm excess}}}
\def\b2t{${\rm B/T}_{\rm I_{814}}$}

\begin{document}

\title{On the Continuous Formation of Field Spheroidal Galaxies in Hierarchical Models of Structure Formation}
\author[Andrew J.~Benson, Richard S.~Ellis, Felipe Menanteau]{Andrew J.~Benson$^1$, Richard S.~Ellis$^1$ \& Felipe Menanteau$^2$ \\
1. California Institute of Technology, MC 105-24, Pasadena, CA 91125, U.S.A. (e-mail: abenson,rse@astro.caltech.edu) \\
2. Carnegie Observatories, 813 Santa Barbara Street, Pasadena, CA 91101, U.S.A. (e-mail: felipe@ociw.edu)}

\maketitle

\begin{abstract}
We re-examine the assembly history of field spheroidals as a
potentially powerful discriminant of galaxy formation models.  Whereas
monolithic collapse and hierarchical, merger-driven, models suggest
radically different histories for these galaxies, neither the
theoretical predictions nor the observational data for field galaxies
have been sufficiently reliable for precise conclusions to be drawn. A
major difficulty in interpreting the observations, reviewed here,
concerns the taxonomic definition of spheroidals in merger-based
models. Using quantitative measures of recent star formation activity
drawn from the internal properties of a sample of distant field
galaxies in the Hubble Deep Fields, we undertake a new analysis to
assess the continuous formation of spheroidal galaxies. Whereas
abundances and redshift distributions of modelled spheroidals are
fairly insensitive to their formation path, we demonstrate that the
distribution and amount of blue light arising from recent mergers
provides a more sensitive approach.  With the limited resolved data
currently available, the rate of mass assembly implied by the observed
colour inhomogeneities is compared to that expected in popular
$\Lambda$-dominated cold dark matter models of structure formation.
These models produce as many highly inhomogeneous spheroidals as
observed, but underpredict the proportion of homogeneous, passive
objects. We conclude that colour inhomogeneities, particularly when
combined with spectroscopic diagnostics for large, representative
samples of field spheroidals, will be a more valuable test of their
physical assembly history than basic source counts and redshift
distributions. Securing such data should be a high priority for the
Advanced Camera for Surveys on Hubble Space Telescope.
\end{abstract}

\section{Introduction}

The star formation and assembly history of field `spheroidals'
(defined here to include both ellipticals and S0s) has been
discussed as one of the most powerful tests of galaxy formation
capable of distinguishing between very different structure
formation possibilities \cite{kauff93,baugh96b,kauff98}.

Traditionally spheroidals were considered to have evolved as
isolated systems following a `monolithic' collapse at high
redshift \cite{eggen62}. Evidence in support of this conjecture
comes from many quarters, including their high central stellar
density and slow rotation, consistent with non-dissipative
collapse at early epochs \cite{efstath83}, and their remarkable
homogeneity at various look-back times in colour-magnitude
\cite{sandage78,bower90,ellis97,stanford98} and fundamental plane
studies \cite{vandokkum01,treu01} of distant clusters.

This picture contrasts with that na\"{\i}vely expected in models where
galaxies assemble hierarchically at a rate governed by the merging of
cold dark matter halos and dynamical friction timescales
\cite{kauff93,baugh96b,cole00}. Distinguishing between both hypotheses
is central to verifying the cold dark matter picture but progress is
hindered by the need to distinguish between star formation histories
and those of mass assembly.

Most of the early work concentrated on ellipticals in rich clusters
where there are significant difficulties of interpretation. The early
epoch of star formation consistent with the colour-magnitude and
fundamental plane studies may still be consistent with hierarchical
models since clusters form from the rare, high density peaks in the
primordial fluctuations within which we can expect accelerated
evolution. Moreover, comparisons of galaxies in individual clusters
selected at various redshifts may be fundamentally affected by the way
in which rich clusters are themselves selected. Conveniences of
observation (e.g. selecting clusters of a given richness, with a
relaxed distribution or of a given X-ray temperature) may bias us
towards the selection of more evolved systems at a given epoch
\cite{kauff96}, thereby giving a misleading impression of a uniformly
old population. Most fundamentally, if galaxies within clusters are
themselves continuously evolving or arriving, it will be difficult to
conduct all-inclusive inventories of a given morphological type as a
function of look-back time.

For the above reasons, attention has now turned towards providing
observational constraints on the evolution of {\it field}
galaxies. The term `field' here includes systems in clusters, of
course, and implies an attempt to be all-inclusive in the study of
the population. The simplest test for evolution in field
spheroidals (and the one which has received most attention) is
concerned with establishing their {\it number density} at high
redshift, either using HST morphologies to isolate spheroidals
\cite{zepf97,schade99,frances99,im01b}, optical-infrared colours to
locate passively-evolving examples whose stars formed at high
redshift \cite{barger99,daddi00,mccarthy00} or some combination of
both \cite{menanteau99}.

We will show in this article that tests which rely on abundance
comparisons at various redshifts or apparent magnitude limits, are
currently inconclusive and, more fundamentally, are poorly-suited to
discriminating between the physical details of the various assembly
hypotheses. In terms of the observations, the currently available
spectroscopic datasets insufficiently sample the redshift domain,
1$<z<$2, where evolution is expected in popular low density models
\cite{daddi00,mccarthy00,firth01}. Comparisons based on counts alone
are reliant on a good knowledge of the shape of the local luminosity
function and its evolution. Without redshift data, there is
considerable ambiguity in the implications \cite{glazebrook95}. In the
case of samples defined morphologically from HST data, there is further
confusion as to the merits of including compact objects which may
possibly be unrelated star-forming galaxies whose abundance seems to
increase significantly at faint magnitudes. Finally, selections based
on red colours alone, even those utilising the new generation of
infrared surveys, permit inclusion of unrelated dusty later-type
galaxies. Colour and morphology are known to poorly correlate at low to
intermediate redshifts \cite{schade99} and beyond $z\simeq$1 where
dusty sources are more likely, the potential for confusion can only
increase. The main drawback of abundance comparisons is that only
subtle changes are expected at the depths currently being probed in low
density universes, thus the significance of claims for or against the
hierarchical and monolithic (or number-conserving pure luminosity
evolution) models remains marginal.

\begin{figure}
\psfig{file=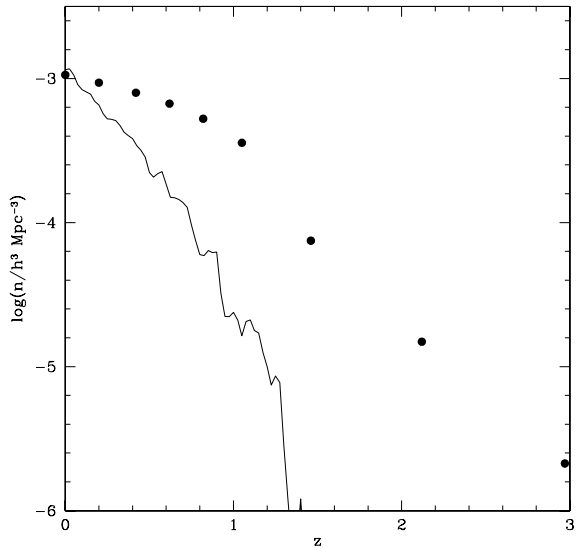,width=80mm}
\caption{Theoretical uncertainties in the evolutionary history of
massive spheroidals: the solid line and circles show the comoving
density of all galaxies with stellar mass greater than
$10^{11}M_\odot$ as a function of redshift in the $\Lambda$CDM models
of \protect\scite{cole00} and \protect\scite{kauff99} respectively.}
\label{fig:KCcompare}
\end{figure}

Significant uncertainties are also present in the theoretical models
used to interpret the observations. In semi-analytic models the build
up of the stellar mass of galaxies is determined using simple
arguments for cooling and star formation rates in a merging hierarchy
of dark matter halos. Even granting this very simple prescription,
there is a surprising uncertainty concerning the expected evolution in
the number density of massive systems. For example, we can consider
the comoving abundance of objects selected by stellar mass as a
function of redshift. In Fig.~\ref{fig:KCcompare} the solid line and
circles show the comoving abundance of galaxies with stellar masses
greater than 10$^{11} M_{\odot}$ in a $\Lambda$CDM cosmology as
predicted by \scite{cole00} and \scite{kauff99} respectively. (Note
that while the basic cosmological parameters, $\Omega_0$ and
$\Lambda_0$ were the same for these two calculations others, including
$\Omega_{\rm b}$, were different. The \scite{cole00} results are taken
from the models used in this paper, while those of \scite{kauff99}
were computed from their publically available galaxy catalogues.)
Although the two models are in reasonable agreement at $z=0$ only
rather modest evolution (about a factor three decline) is predicted to
$z\simeq$1 by \scite{kauff99}, whereas much stronger evolution is
expected according to the model of \scite{cole00}.

This discrepancy in the rate of evolution predicted by two similar
models is very worrying, although it should be remembered that the
galaxy stellar mass function is steep in this region. Hence even minor
differences in the rate at which galaxy masses evolve in the two
models could lead to large differences in abundance. For example, the
factor of 10 difference at $z=1$ might be resolved by considering
galaxies only a factor of 2 less massive in the Cole et
al. model. Nevertheless, it is evidently crucial to understand these
differences before strong tests of the hierarchical hypothesis can be
made.

A recent analysis of spheroidals in both Hubble Deep Fields (HDF) by
\scite{felipe} offers an interesting opportunity to bypass many of
the difficulties discussed above. Resolved ${\rm V}-{\rm I}$ colour
data for the HDF spheroidals reveals the presence of blue cores in a
significant subset and preliminary spectroscopy (c.f. Ellis 2000)
confirms that recent star formation is the likely cause. Together with
a careful appraisal of the way in which spheroidal galaxies are
defined in semi-analytic models, further analysis of this dataset
forms the basis of the present paper. Instead of making absolute count
comparisons, we use the distribution of colour inhomogeneities to
determine the fraction of spheroids still assembling. We compare this
with hierarchical predictions taking into account the selection
technique and uncertainties arising from the taxonomical definition of
a recognisable spheroidal galaxy. This type of comparison, in which
the information of interest is encoded in a complex way in the data
and important selection criteria exist, is one of the key areas where
semi-analytic models excel. Most significantly, we demonstrate that
such an analysis, based on post-merger relaxed systems with residual
star formation, is a more direct test of the pertinent assembly
details than traditional techniques based on counts or luminosities
alone.

A plan of the paper follows. In \S\ref{sec:define} we discuss the
various ways in which spheroidals can be defined in hierarchical
models, introducing a definition that incorporates a timescale for
relaxation to a spheroidal form. We test predictions utilising these
definitions against published morphological number counts
\cite{glazebrook95,driver95dummy,frances99} and redshift distributions
\cite{brinchmann00,menanteau99}. In \S\ref{sec:implic} we revisit the
observational analysis of Menanteau et al., and compare theory with
observation in the light of both observational and theoretical
uncertainties. We summarise our conclusions in \S\ref{sec:discuss}.

\section{Spheroidal Galaxies in Hierarchical Models}
\label{sec:define}

\subsection{Semi-Analytic Modelling and Morphological Evolution}

First we discuss the basic processes which govern the mass assembly
and star formation histories of spheroidals in hierarchical
models. Specifically, we will utilise the galaxy formation model
described by \scite{cole00} to which the reader is referred for a full
description. Here we summarise briefly the salient features. A small
number of model parameters are crucial in determining the properties
of the spheroids in which we are interested as described below. Cole
et~al. describe how these model parameters may be constrained by a set
of $z=0$ observational data. We retain the same parameter values which
Cole et~al. found produced the best agreement with the local data. As
such, the model results presented in this work are true
predictions. In \S\ref{sec:implic} we will consider the effects of
changing some of these parameters (and also in altering the basic
mechanisms by which spheroidal galaxies are made) as a way to
understand the model results and to test their robustness and also to
ascertain the prospects of our approach for directly constraining the
formation history of spheroidals. It must be kept in mind however,
that once a parameter value is changed the model is unlikely to remain
a good fit to the local data and so any such results cannot be
considered to be predictions of a realistic hierarchical model.

In this model, galaxies form continuously from cooling gas inside a
dark matter halo. Gas settles into a disk configuration within which
stars form quiescently. Spheroidal galaxies are produced by a major
merger of two pre-existing galaxies of comparable mass (note that the
pre-existing galaxies may be disks, spheroids or a mixture of both,
and that disks may contain both stars and gas), which occurs due to
dynamical friction (the timescale of which is determined by a
parameter of the model, $f_{\rm df}$, which is set to $1$ unless
otherwise stated) dissipating energy from the galaxies' orbits. Such a
merger disrupts the galaxies leaving a spheroidal remnant, and also
triggers any cold gas present to undergo a burst of star formation
with an exponentially decreasing star formation rate. The spheroidal
so formed may accrete gas through cooling at a later time and thereby
grow a new disk. Model galaxies can thus migrate either way along the
Hubble sequence, and may change their morphologies several times
during their lifetimes.

Only a major merger (defined as one in which $M_1/M_2\ge f_{\rm
ellip}$, where $M_1$ and $M_2$ are the masses of the merging
components with $M_1\leq M_2$ and $f_{\rm ellip}$ is a parameter of
the model) triggers a burst of star formation.  The timescale of the
burst is $\tau_{\rm burst}=\epsilon_{\star,{\rm burst}}^{-1} \tau_{\rm
dyn}$, where $\tau_{\rm dyn}$ is the dynamical time of the newly
formed spheroid and $\epsilon_{\star,{\rm burst}}$ is a free
parameter.

\scite{cole00} chose $f_{\rm ellip}=0.3$, as suggested by numerical
simulations \cite{walker96,barnes98} which show that the merging
galaxy mass ratio must be in the range $0.3$ to $1.0$ to destroy the
progenitor components and produce a spheroidal remnant. The value of
$\epsilon_{\star,{\rm burst}}$ is best constrained by the luminosities
of bursting galaxies since these are directly related to the star
formation rate in the burst. To accurately predict the luminosities of
starbursts a detailed treatment of the absorption and reprocessing of
light by dust is needed. Such an analysis has been carried out by
\scite{granato00}, who found that $\epsilon_{\star,{\rm burst}}=0.5$
provided the best fit to the $z=0$, 60$\mu$m galaxy luminosity
function (in which we see light absorbed and re-emitted by dust in
starbursts) so we will adopt this value as default in this work.

Spheroidals may also form via the gradual accretion of smaller
galaxies (minor mergers). In the case of a minor merger ($M_1/M_2 <
f_{\rm ellip}$) we assume no burst of star formation is
involved. Stars from the smaller galaxy are added to the bulge of the
larger galaxy, which may then become dominant, while any residual gas
is added to its disk.

\subsection{Model Comparisons based on Number Counts and Redshift Distributions}
\label{sec:compare}

In this section we are concerned with exploring those parameters
central to the taxonomical definition of a morphological
spheroidal. The majority of previous studies using semi-analytic
models to study the properties of galaxies selected by their
morphological type \cite{kauff96,baugh96dummy,kauff98} have used the
bulge-to-total ratio, B/T, usually (as here) measured in terms of
luminosity. (Specifically, in this work we use the dust-extinguished
I$_{814}$-band luminosity to define the bulge-to-total ratio, \b2t.)
\scite{baugh96a} also considered the time since the most recent major
merger for each galaxy to be a defining property, placing galaxies
whose last major merger occurred less than 1 Gyr earlier into the
Irr/Pec category. Accordingly, spheroidals in the \scite{baugh96a}
scheme jointly satisfy B/T$>0.4$ and have experienced no major merger
in the last 1 Gyr.

The key issue, central to a successful interpretation of the resolved
colour data, is that spheroidals can appear to be dynamically relaxed
while showing evidence of recent star formation (e.g. in their
internal colours). Instead of adopting a fixed relaxation period to
define a semi-analytical spheroidal, we include a criterion based on
the dynamical timescale of the remnant. After a major merger, the
remnant begins to relax from the inside out in a time comparable to
the local dynamical time \cite{barnes92}. For each remnant we
calculate the dynamical time at the half-mass radius. Any object which
experienced a major merger less than $N_{\rm relax}$ dynamical times
ago is considered to be an Irr/Pec galaxy, irrespective of its
B/T. $N_{\rm relax}$ is a parameter of the model which we expect to be
of order unity.  For the sample of galaxies considered in this work,
spheroidal dynamical times are typically a few times $10^7$ years,
i.e. much shorter than 1~Gyr. Crucially, for small values of $N_{\rm
relax}$, a merger remnant can be classed as relaxed well before the
distinctive blue colour of recent star formation activity has
declined.

We summarise our scheme for defining a sample of morphological spheroidals 
in Table~\ref{tb:scheme}. Since in this work we are
primarily interested in the E/S0 class only the parameters B/T$_{\rm
S}$ and $N_{\rm relax}$ need to be constrained. As such, we will not
attempt to constrain the values of the other parameters in
Table~\ref{tb:scheme}. Although \scite{cole00} briefly considered the
morphological mix of bright galaxies, they did not explore in detail
the sensitivity of their parameter choices to data on morphologically
selected faint galaxies. As noted by \scite{baugh96b} we have some
guidance as to the value of B/T$_{\rm S}$ from comparisons of galaxy
T-types with their bulge-to-total ratio obtained from bulge/disk
decomposition studies. Such comparisons suggest B/T$_{\rm S}\approx
0.4$, but with significant uncertainty. Keeping this figure in mind,
we now explore what constraints can be set on B/T$_{\rm S}$ and
$N_{\rm relax}$ from the available observational data, specifically
morphologically selected number counts and redshift distributions.

\begin{table}
\caption{Morphological classification scheme. Galaxies are classified
on the basis of their bulge-to-total ratio in dust-extinguished I-band
light, B/T$_{\rm I_{814}}$ (column 1), and the ratio of the time since
the last burst of star formation to the dynamical time of the
spheroid, $t_{\rm burst}/t_{\rm dyn}$ (column 2). The morphological
class assigned to each region of this two-parameter space is given in
column 3.}
\label{tb:scheme}
\begin{tabular}{r@{}c@{}llc}
 & B/T$_{\rm I814}$ & & $t_{\rm burst}/t_{\rm dyn}$ & Morphological class \\
\hline
$0$& $\leq {\rm B/T} \leq$& ${\rm B/T_{Irr}}$ & $> N_{\rm relax}$ & Irr \\
${\rm B/T_{Irr}}$ &$< {\rm B/T} \leq$& ${\rm B/T_{S}}$ & $> N_{\rm relax}$ & S \\
${\rm B/T_{S}}$& $< {\rm B/T} \leq$& 1 & $> N_{\rm relax}$ & E/S0 \\
$0$& $\leq {\rm B/T} \leq$& $1$ & $\leq N_{\rm relax}$ & Pec
\end{tabular}
\end{table}

\begin{table}
\caption{Parameters of the relevant morphological classification
schemes (but which do not affecting the merging and formation
histories of model galaxies).}
\label{tb:stmodparams}
\begin{center}
\begin{tabular}{lccc}
 & \multicolumn{3}{c}{Parameter set} \\
\cline{2-4}
Parameter & Default & Alt. B/T$_{\rm S}$ & Alt. $N_{\rm relax}$ \\
\hline
${\rm B/T}_{\rm S}$ & 0.44 & 0.30 & 0.44 \\
$N_{\rm relax}$ & 3 & 3 & 30
\end{tabular}
\end{center}
\end{table}

\begin{figure}
\psfig{file=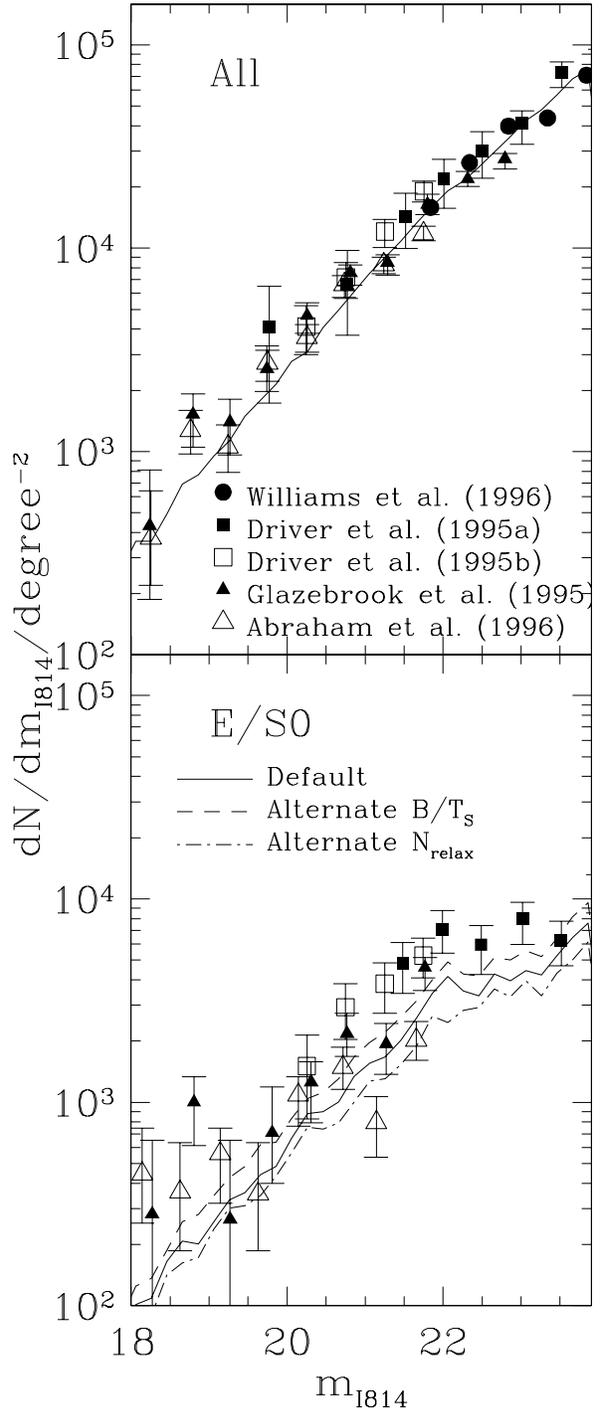,width=80mm,bbllx=0mm,bblly=15mm,bburx=105mm,bbury=265mm,clip=}
\caption{Differential number counts in the I$_{\rm 814}$ pass-band as
a function of morphological type.  Panels show the counts for all
galaxies (top) and E/S0 galaxies (bottom). Points show a compilation
of observational determinations (as indicated in the upper panel)
while lines show the model results (solid line --- default parameters;
dashed line --- alternative B/T$_{\rm S}$; dot-dashed line ---
alternative $N_{\rm relax}$).}
\label{fig:counts}
\end{figure}

Number counts are only weakly dependent on the assumed model
parameters. Figure~\ref{fig:counts} shows the E/S0 and total number
counts in the $I_{814}$ band\footnote{We use Vega-based magnitudes
throughout this paper.}; clearly the baseline model (solid lines)
matches these data very well. The key question is the extent to which
the morphologically-classified counts depend on our parameters
B/T$_{\rm S}$ and $N_{\rm relax}$. Both the S and Irr/Pec number
counts (not shown) are relatively insensitive to the value of $N_{\rm
relax}$ and thus constrain the parameter B/T$_{\rm S}$ to be $\lsim
0.44$ (in good agreement with the figure expected from a comparison to
T-types). A value as low as B/T$_{\rm S}=0.30$ remains consistent with
the counts, while lower values begin to underpredict the number
of spirals. The predicted E/S0 counts are quite insensitive to $N_{\rm
relax}$ for $N_{\rm relax}\lsim 10$. The dot-dashed line in
Fig.~\ref{fig:counts} shows the results for $N_{\rm relax}=30$, which
begins to deplete the number of E/S0 galaxies below that
observed. While the counts suggest a reasonably low value of $N_{\rm
relax}$, they do not provide a strong constraint.
This insensitivity might be perceived as an advantage; a detail in
the assembly history is irrelevant in interpreting an observable.
However, as the relaxation timescale is central to the rate at
which systems become recognisable spheroidals, this would be false
security, particularly in diagnosing systems at high redshift.

In the light of the above constraints, in the remainder of this paper
we will consider three sets of values for the parameters B/T$_{\rm S}$
and $N_{\rm relax}$ as listed in Table~\ref{tb:stmodparams}. Note that
these parameters determine only the morphological classification
scheme and, as such, we would ideally constrain them accurately from
morphologically selected number counts, redshift data etc. Later, we
will consider altering other parameters of the model which directly
alter the formation and merging histories of model galaxies. It is
these formation and merging rates which we aim to constrain through
the methods described in this paper.

In the upper two panels of Fig.~\ref{fig:dndz} we compare the redshift
distribution of model galaxies in magnitude limited samples with the
recent HST-based redshifts surveys as summarised by
\scite{brinchmann00}. These encompass the initial survey of
\scite{schade99} enlarged to include a total of 54 spheroidals
spectroscopically-complete to I$_{814}=22$. Additional data to a
somewhat brighter magnitude limit is in principle available from the
survey of \scite{im01a} and \scite{im01b}. However, we chose not to
include this sample as the spheroidals were defined according to
symmetry and colour, i.e. differently than in Figure~\ref{fig:counts}.
\scite{im01b} discuss the uncertainties of their definition which
affect the absolute abundances at the $\simeq\pm$25\% level. As
expected, the proportion of high redshift spheroidals is sensitive to
$N_{\rm relax}$, with larger values producing fewer examples (since
more galaxies are placed into the Irr/Pec class). Unfortunately,
although the range of redshifts spanned by the model galaxies is
comparable to that seen in the data, the detailed correspondence
between theory and observations is not particularly good for the total
sample (upper panel), a point discussed by \scite{baugh96a}.

The lower panel of Fig.~\ref{fig:dndz} shows the deeper (I$_{814}<24$)
E/S0 sample of \scite{menanteau99}, which is employed extensively in
this paper. Barring one discrepant point, the model is in very good
agreement with this redshift distribution. The differences between the
model predictions for the three parameter sets illustrate again how
well these parameters could be constrained by larger datasets.

\begin{figure}
\psfig{file=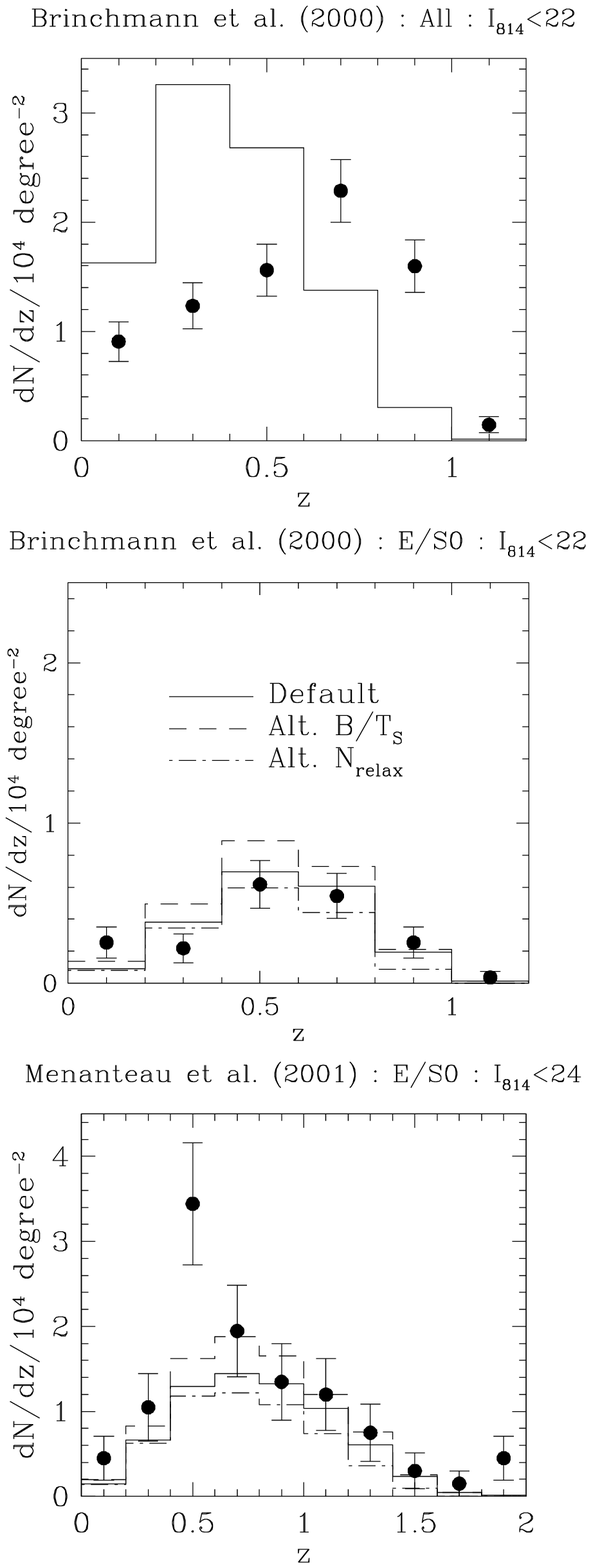,width=80mm,bbllx=10mm,bblly=15mm,bburx=105mm,bbury=270mm,clip=}
\caption{}
\end{figure}

\begin{figure}
\addtocounter{figure}{-1}
\caption{\emph{(cont.)} The three panels show faint galaxy redshift
distributions, with observational data indicated by filled circles and
model predictions shown as histograms. In the upper panel we show the
redshift distribution of galaxies of all morphological types brighter
than I$_{814}=22$ from \protect\scite{brinchmann00}, while the middle
panel shows the redshift distribution of spheroidal galaxies from the
same sample. The lower panel shows the redshift distribution of
I$_{814}<24$ spheroidals from \protect\scite{menanteau99}. In the
lower two panels, model results are shown for standard (solid
histogram), alternative B/T$_{\rm S}$ (dashed histogram) and
alternative $N_{\rm relax}$ (dot-dashed histogram) parameters.}
\label{fig:dndz}
\end{figure}

While our baseline semi-analytical model, in which spheroidals are
formed by the mergers of two pre-existing galaxies, produces some of
the observed trends in this population, at a quantitative level it
disagrees with the present redshift data. More relevant at this stage,
however, is the poor sensitivity of such observables to key physical
features of the hierarchical model. Even with exquisite data at these
magnitude limits, we can expect it will be hard to conclusively
determine whether the hierarchical model is correct given the freedoms
in the morphological definitions introduced in Table 1. Therefore, we
propose to use these observational data to constrain the morphological
classification parameters of Table~\ref{tb:scheme} and then use other
data, such as that described in this paper, to constrain the formation
and merger rates of spheroidals.

\subsection{Quantifying Recent Activity Using Colour-Based Properties}
\label{sec:quant}

In the present section we demonstrate that the colour inhomogeneities
discovered for HDF spheroidals \cite{felipe} offer a much more direct
probe of the hierarchical hypothesis than the number count and
redshift distributions discussed in \S\ref{sec:compare}.  Not only do
the {\it internal} colour variations seen in the HDF offer an
important way to separate models based on isolated passive evolution
following monolithic collapse from those invoking continuous
hierarchical assembly, but they also provide the first opportunity for
a quantitative measure of the assembly rate in the latter
case. Importantly, \scite{felipe} find that around one third of
spheroidals in their sample show marked inhomogeneities in their
internal colours, and that in the majority of these cases the
inhomogeneity is characterised by a blue core in the galaxy colour
profile. The observations therefore require that blue light be more
centrally concentrated than red light. Here we will assume that these
blue cores arise due to recent bursts of star formation.

For present purposes, where we are concerned only with V and I
photometry, the major merger event which forms each elliptical in the
model can be characterised by two physical quantities: (i) the
fraction of stars formed in the associated burst relative to the final
stellar mass of the spheroidal remnant, $f_{\rm burst}$, and (ii) the
time since the merger, $t_{\rm burst}$\footnote{The metallicities of
pre-existing and newly-formed stars will also have some effect on the
colours, an effect correctly accounted for in the model, but is of
secondary importance to the present discussion.}. The first determines
how the colour of the remnant evolves with time, and which can in
principle be constrained by the resolved HDF data, and the second
determines at what point in that evolution the galaxy is
observed. Ideally we would like to measure these two parameters
directly from the observational data as this would allow us to
reconstruct the rate of spheroid formation as a function of
redshift. However, with the present photometric data and relatively
small sample size such an approach is not warranted (spectroscopic
measurements will provide far better constraints on these
parameters). Instead we focus on the much simpler objective of
assessing the compatibility of hierarchical formation with the present
data. However, we will give predictions for the variation of these key
parameters with redshift which may then be tested by forthcoming
observations.

\begin{figure*}
\begin{tabular}{cc}
\psfig{file=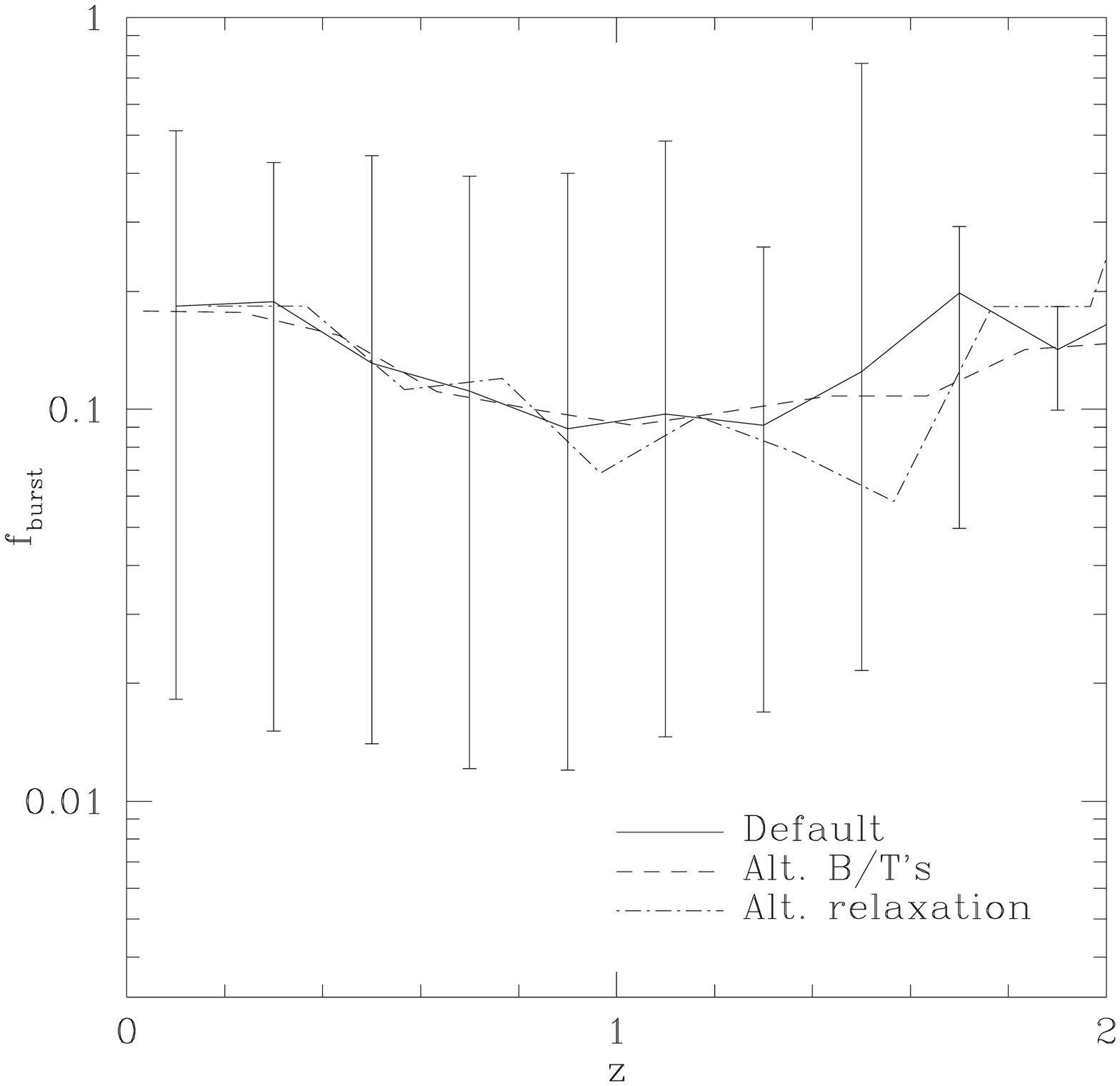,width=80mm} & \psfig{file=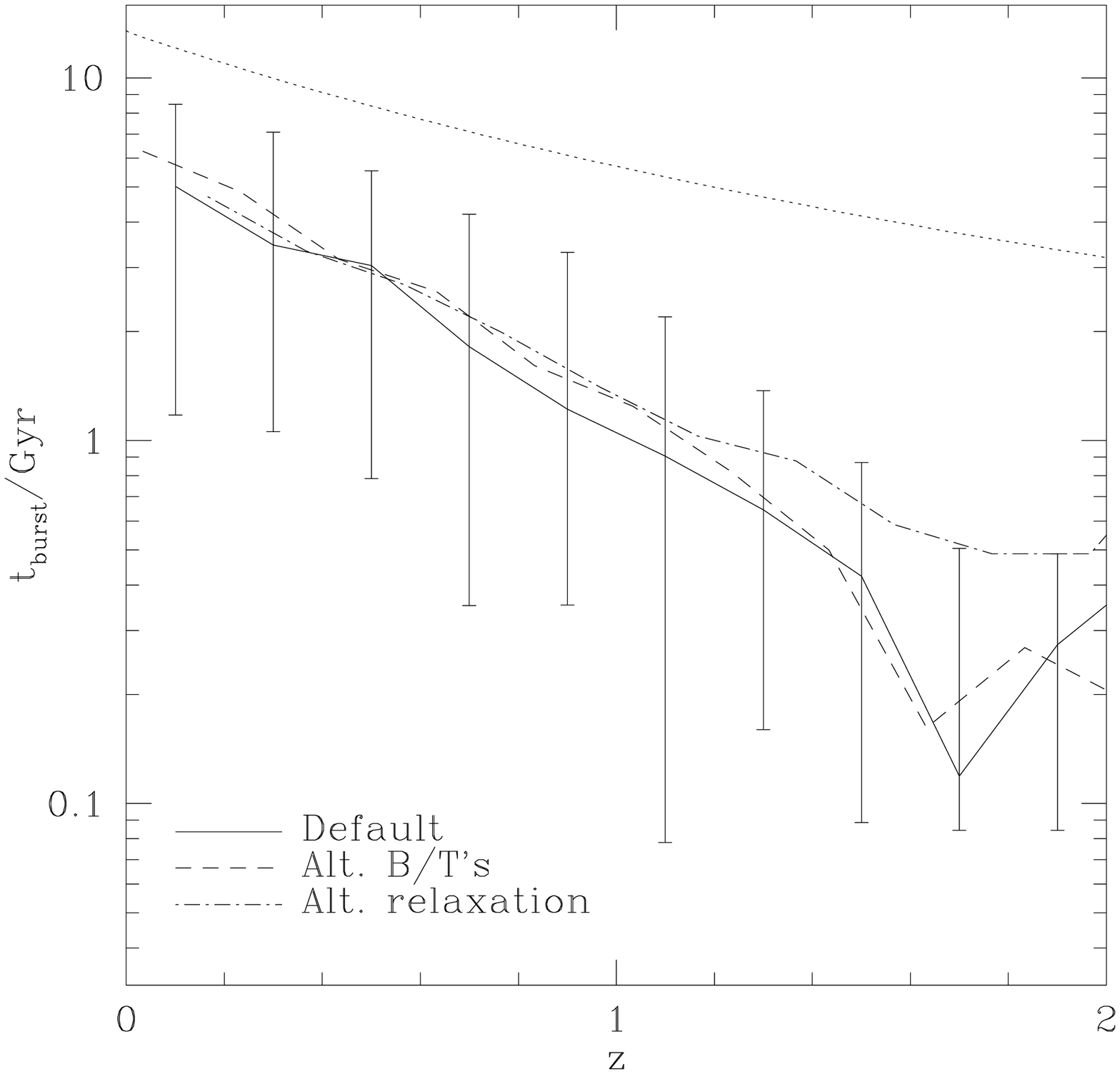,width=80mm}
\end{tabular}
\caption{The variation with redshift of the two parameters defining a
merger event, $f_{\rm burst}$ and $t_{\rm burst}$ (left and right-hand
panels respectively). Lines show the median of the model relation with
error bars indicating the 10\% and 90\% intervals. Solid lines show
results for our default parameters, dashed and dot-dashed for the
alternative B/T$_{\rm S}$ and $N_{\rm relax}$ respectively. For
clarity, error bars are shown only on the solid lines. The 10\% and
90\% intervals for the other models are comparable to the default
model with the exception that the 10\% intervals of $t_{\rm burst}$
for the alternative $N_{\rm relax}$ model occur at significantly
larger $t_{\rm burst}$.}
\label{fig:real}
\end{figure*}

In Fig.~\ref{fig:real} we show how the two physical quantities vary
with redshift for a magnitude-limited sample of model galaxies in a
$\Lambda$CDM cosmology. In this and subsequent figures, we will
consider spheroidal galaxies brighter than $I_{814}=24$ in order to
match the HDF sample used by \scite{felipe}. Unless otherwise stated
all model results refer to our default parameters (refer to
Table~\ref{tb:stmodparams}).

There is little variation in $f_{\rm burst}$ with redshift, but a
significant dispersion at fixed $z$. The median burst typically
involves 10-20\% of the total spheroidal mass. The lack of any trend
with redshift is the result of two counteracting effects. Firstly,
$f_{\rm burst}$ shows a strong correlation with the mass of the
spheroidal galaxy, such that lower mass galaxies tend to have higher
values of $f_{\rm burst}$. This trend arises because more massive
ellipticals tend to form from more massive disk progenitors
\scite{kauff98} and, in the model of \scite{cole00}, more massive
disks tend to have smaller gas fractions (at least if we consider only
central galaxies, which are always involved in the formation of an
elliptical), a consequence of the variation of star formation rate
with disk mass.

The apparent magnitude limit of the sample prevents us from seeing
the least massive (highest $f_{\rm burst}$) galaxies at high
redshifts. However, at high redshift there are also fewer of the
most massive (lowest $f_{\rm burst}$) galaxies (since these have
not had time to form).

There is a strong trend of $t_{\rm burst}$ with redshift however. This
quantity is bounded by the age of the universe at given $z$ (since no
galaxy can have experienced a burst which began earlier than
$t$=0). The dotted line in Fig.~\ref{fig:real} shows the age of the
universe as a function of redshift. The evolution in the median value
of $t_{\rm burst}$ is much more rapid than this. This is a consequence
of higher rates of merging at high redshift combined with the fact
that, at high redshift, a galaxy is seen only if it is intrinsically
luminous and so we preferentially select galaxies which have recently
experienced a burst of star formation, these being more luminous than
quiescent examples (particularly in view of the k-correction at
$z\approx 1.5$).

In conclusion the model predicts that for the magnitude limited sample
considered here, a strong correlation between $t_{\rm burst}$ and
redshift should exist, such that higher redshift spheroidals have
typically experienced a burst more recently.

We now describe how we measure the degree of recent activity in a
spheroidal galaxy from the available photometric data (both real
observational data and mock images of model galaxies). \scite{felipe}
proposed a statistic \index\ which measures the degree of
inhomogeneity in the internal colours defined as:

\begin{equation}
\delta({\rm V}-{\rm I})=2N{\sum (x_i-\bar{x}_i)^2 S(x_i){\rm SNR}(x_i)\over \sum S(x_i){\rm SNR}(x_i)},
\end{equation}

where $x_i$ is the colour of the $i^{\rm th}$ pixel, ${\rm SNR}(x_i)$
is the signal to noise ratio, $S(x_i)$ is a step function which
includes only those pixels with ${\rm SNR}(x_i)>1.3$ in the sum and
$N$ is an arbitrary constant which was taken to be $N=3$. The sum is
taken over all pixels brighter than ${\rm V}_{606}=25.5$mag
arcsec$^{-2}$. The value of \index\ obtained for any single galaxy
will depend on the point spread function (PSF) of the observing
instrument, since substantial smoothing will tend to homogenise the
internal colours. Although wavelength dependent PSFs, such as occur in
the WFPC2, can create artificial colour inhomogeneities. \scite{felipe}
found this to be a negligible source of inhomogeneity for the HDF
spheroidals except for a few galaxies with highly concentrated surface
brightness profiles.

\begin{figure}
\psfig{file=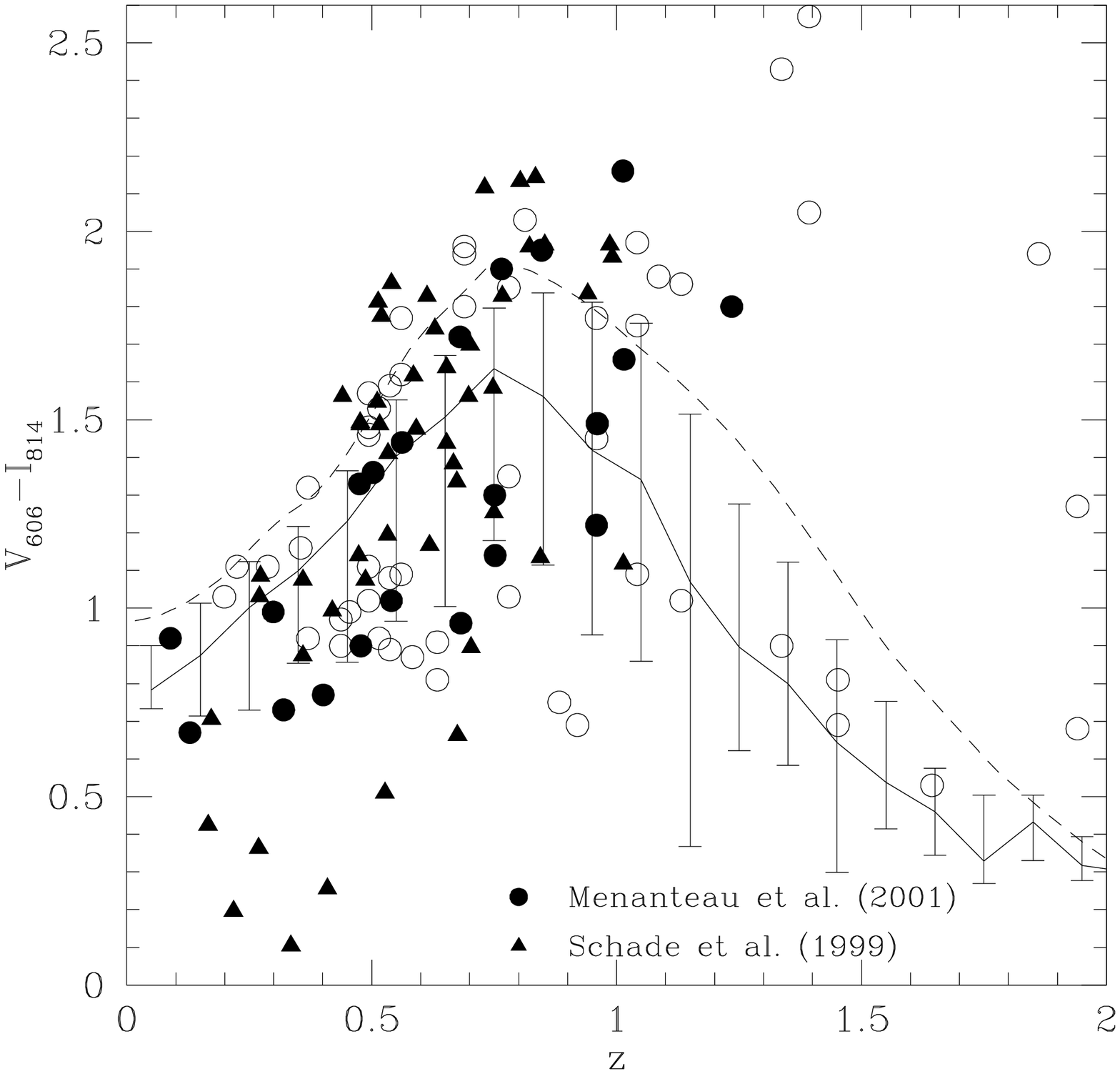,width=80mm} \caption{Integrated V$-$I colour as
a function of redshift. Circles show data from the HDF North and South
with solid points indicating spectroscopic redshifts and open points
photometric redshifts. Triangles show brighter data from the survey of
\protect\scite{schade99} (with morphologies classified visually). The
median of the model distribution is shown by the solid line, with
error bars indicating 10\% and 90\% intervals of the distribution.
The dashed line indicates the colour of a galaxy formed at $z=3$ with
an exponentially decaying star formation rate with e-folding time of
1~Gyr and with Solar metallicity.}
\label{fig:passive}
\end{figure}

As an alternative diagnostic we consider the difference in
integrated colour compared to that expected at the galaxy redshift
for an old, passively evolving population, $({\rm V}-{\rm I})_{\rm
passive}(z)$. The latter case is defined here as a system which
formed at $z_F=3$ with an exponentially decaying star formation
rate with e-folding time of 1 Gyr and with Solar metallicity. In
Fig.~\ref{fig:passive} we show the integrated ${\rm V}-{\rm I}$
colours of observed and model spheroidal galaxies as a function of
redshift and compare these to $({\rm V}-{\rm I})_{\rm
passive}(z)$. We plot data from the HDF sample of \scite{felipe}
and also from the brighter (I$_{\rm AB}<22$) sample of
\scite{schade99} to demonstrate that both datasets display similar
trends. Note that while both model and data broadly trace out the
shape of the passive curve, both reveal many galaxies that are
significantly bluer. This is particularly noticeable for the model
at $z\gsim 1$ where the median colour is $\sim 0.3$ magnitudes
bluer than the passive prediction.

Interestingly, the HDF data also contains a few very {\it red} objects
(i.e. redder than the passively evolving track shown in
Fig.~\ref{fig:passive}, and in fact redder than passively evolving
tracks for galaxies formed even at higher $z_F$). The most anomalous
sources are ones with photometric redshifts above unity. The red
colours could arise as a consequence of an overestimated photometric
redshift although two of these galaxies would be extremely red
whatever redshift they were placed at. Certainly,
Fig.~\ref{fig:passive} suggests that our present model does not
produce spheroidal galaxies as red as those observed in the HDF.

We define the colour excess:
\begin{equation}
\excess = ({\rm V}-{\rm I}) - ({\rm V}-{\rm I})_{\rm passive}(z).
\end{equation}
While this simple measure has the desirable property of being largely
independent of the PSF it makes no use of the resolved internal
structure, and furthermore is sensitive to errors in the redshift of
the galaxy (which is particularly important for the galaxies with only
photometric redshifts). Nevertheless, it is interesting to compare
this measure with the \index\ index.

To determine the sensitivity of both statistics for the model galaxies
we construct a mock image of each galaxy (including noise which makes
a small contribution to \index), smoothed with a Gaussian PSF of FWHM
0.126 and 0.143 arcseconds in the V and I-bands respectively to mimic
the PSF of the WFPC-2 \cite{casertano00}. Both statistics described
above are then measured as for a real galaxy. For the purposes of
constructing mock images each galaxy is assumed to consist of three
components, a disk, a spheroidal remnant and a recent burst. Although
a galaxy may have experienced more than one burst in its history we
treat only the stars formed in the most recent burst as a separate
component, with stars which were formed in earlier bursts assumed to
be uniformly mixed in the spheroidal component. Our results are
insensitive to this assumption, since placing the light from stars
formed in all the bursts in a separate component does not make any
significant difference to the statistics we consider.) Disks are
modelled with an exponential surface brightness profile and random
inclination. \scite{hernquist92} has shown that mergers between purely
stellar systems, specifically bulge-less galaxies, result in a remnant
which has a surface brightness profile close to an $r^{1/4}$-law, but
with a core of constant surface brightness. A simple parameterisation
which displays qualitatively similar behaviour is
\begin{equation}
\Sigma(r) = \Sigma_0 \exp \left( -7.67 \left[ \left( [r/r_{\rm
e}]^2 + c^2 \right)^{1/8} - 1 \right] \right), \label{eq:hq}
\end{equation}
where $r$ is the radius, $r_{\rm e}$ the effective radius and $c$ is a
core radius in units of $r_{\rm e}$. We will refer to this form as the
``Hernquist profile'' hereafter. This expression is not proposed as an
accurate fit to the \scite{hernquist92} results, but has the desirable
properties of being analytically integrable and able to produce
remnants with any desired core size. We distribute the mass of the
pre-existing stars with this surface density profile. Gas in mergers
is generally driven to the centre of the remnant where it presumably
undergoes a burst of star formation \cite{barnes96}. The stars formed
in the recent burst are therefore distributed in such a way that the
total surface mass density (i.e. the sum of burst and pre-existing
components) follows an $r^{1/4}$-law (i.e. as eqn.~\ref{eq:hq} with
$c=0$). The total mass in the Hernquist profile is given by
$M(c)=\int_0^\infty 2 \pi r \Sigma(r) {\rm d}r$. Requiring that
$M(c)/M(0)=M_{\rm pre}/(M_{\rm burst}+M_{\rm pre})$, where $M_{\rm
burst}$ is the mass of the stars formed in the burst and $M_{\rm pre}$
is the pre-existing stellar mass, allows us to solve for $c$. The
value of $r_{\rm e}$ is determined from the half-mass radius of the
remnant as calculated by the galaxy formation model \cite{cole00}.
The light of each component is then distributed as the mass assuming a
constant mass-to-light ratio. With this construction the spheroid
would eventually come to resemble an $r^{1/4}$-law at late times when
the stars of both components were old (ignoring any differences in
metallicity between the two populations).

In cases where the recent burst accounts for a large fraction of the
total mass of the spheroid the core radius, $c$, can become very
large. No simulations of mergers between such gas-rich disks have been
carried out (but see \pcite{somerville01}), so the above form for the
pre-existing and burst star populations may not be relevant in such
cases. We therefore consider an alternative case in which both
pre-existing and burst stars are distributed as $r^{1/4}$-law
profiles, but with the burst stars having a much smaller effective
radius (one tenth of the spheroid radius).

\section{Analysing Recent Activity in HDF Ellipticals}
\label{sec:implic}

We now compare the distributions of model and observed HDF
spheroidal galaxies using the two statistics described in
\S\ref{sec:quant}.

\begin{figure*}
\psfig{file=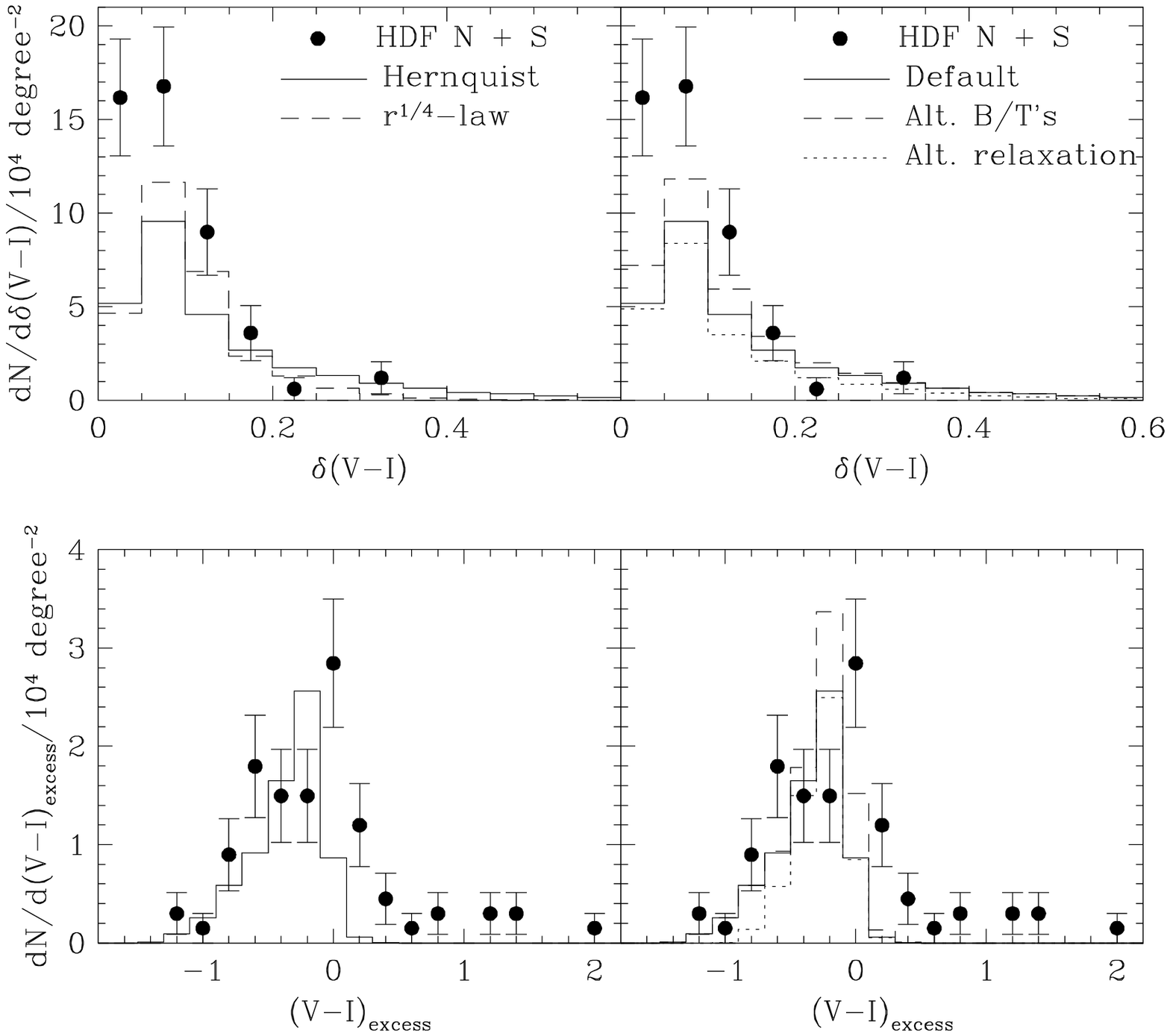,width=160mm,bbllx=0mm,bblly=100mm,bburx=190mm,bbury=270mm,clip=}
\caption{Differential distributions of both statistics discussed in
\S\protect\ref{sec:quant} for E/S0 galaxies brighter than I$_{\rm
814}=24$. Circles show observational data from the combined HDF North
\& South. Histograms show model results. In the left-hand column solid
histograms are for the Hernquist profile, while dashed histograms use
an $r^{1/4}$-law profile. In the right-hand column solid lines are for
the default parameters, dashed for for alternative B/T$_{\rm S}$ and
dotted for alternative $N_{\rm relax}$.}
\label{fig:statistics}
\end{figure*}

The lower left-hand panel of Fig.~\ref{fig:statistics} shows
differential distributions of each statistic. For \excess\ the model
produces a comparable distribution to the data for blue objects
($\excess \lsim0$), but cannot account for the red objects
($\excess\gsim0$) as was expected on the basis of
Fig.~\ref{fig:passive}. This quantity is independent of the surface
brightness profiles we ascribe to model galaxies, but does depend on
the values of the parameters B/T$_{\rm S}$ and $N_{\rm relax}$. In the
lower right-hand panel of Fig.~\ref{fig:statistics} we demonstrate the
effects of using the alternative parameter values given in
Table~\ref{tb:stmodparams}. Using B/T$_{\rm S}=0.30$ instead of 0.44
produces substantially more galaxies just bluewards of $\excess=0$
(this alternative morphological definition tends to include several
galaxies which experienced their last burst much longer than 1Gyr ago
and whose only star formation is occurring in their disks), without
affecting the rest of the distribution. Increasing $N_{\rm relax}$
from 3 to 30 significantly reduces the number of very blue model
galaxies. With the larger value of $N_{\rm relax}$ these extremely
blue galaxies, which have typically experienced a major merger in the
very recent past, are rejected from the E/S0 class. With the present
data we can infer $N_{\rm relax}\lsim 10$. Note that neither of these
parameter alterations helps reconcile the observed and predicted
distributions of red galaxies.

For \index\ we show two model lines in the upper left-hand panel, the
solid one using the Hernquist profile and the dashed one using an
$r^{1/4}$-law for both burst and pre-existing stellar components. The
choice of profile does not drastically alter the width or shape of the
distribution. Using $r^{1/4}$-law profiles produces somewhat more
galaxies at low values of \index\ and does not produce the tail to
high values which occurs with the Hernquist profile. We remind the
reader that while the Hernquist profile has some theoretical
justification, the $r^{1/4}$-law profile has none. The purpose of
considering this profile is to demonstrate that the model
predictions are not overly sensitive to this choice.

In the upper left-hand panel we again see that the model distribution
is in reasonable agreement with the data for high values of \index,
but appears to predict too few galaxies with low \index. In the upper
right-hand panel we consider the effects of changing model parameters
on the predictions for \index. Changing B/T$_{\rm S}$ from 0.44 to
0.30 produces more model galaxies at low \index, thereby improving the
agreement between model and data (although some differences remain),
while increasing $N_{\rm relax}$ to 30 depletes the number of galaxies
at all \index, but particularly so at high \index\ since these
galaxies are those which have experienced a major merger in the very
recent past. However, even with this large value of $N_{\rm relax}$
(which is marginally ruled out by the morphologically selected counts
of Fig.~\ref{fig:counts}) the model still provides a reasonable match
to the high \index\ data.

\begin{figure*}
\psfig{file=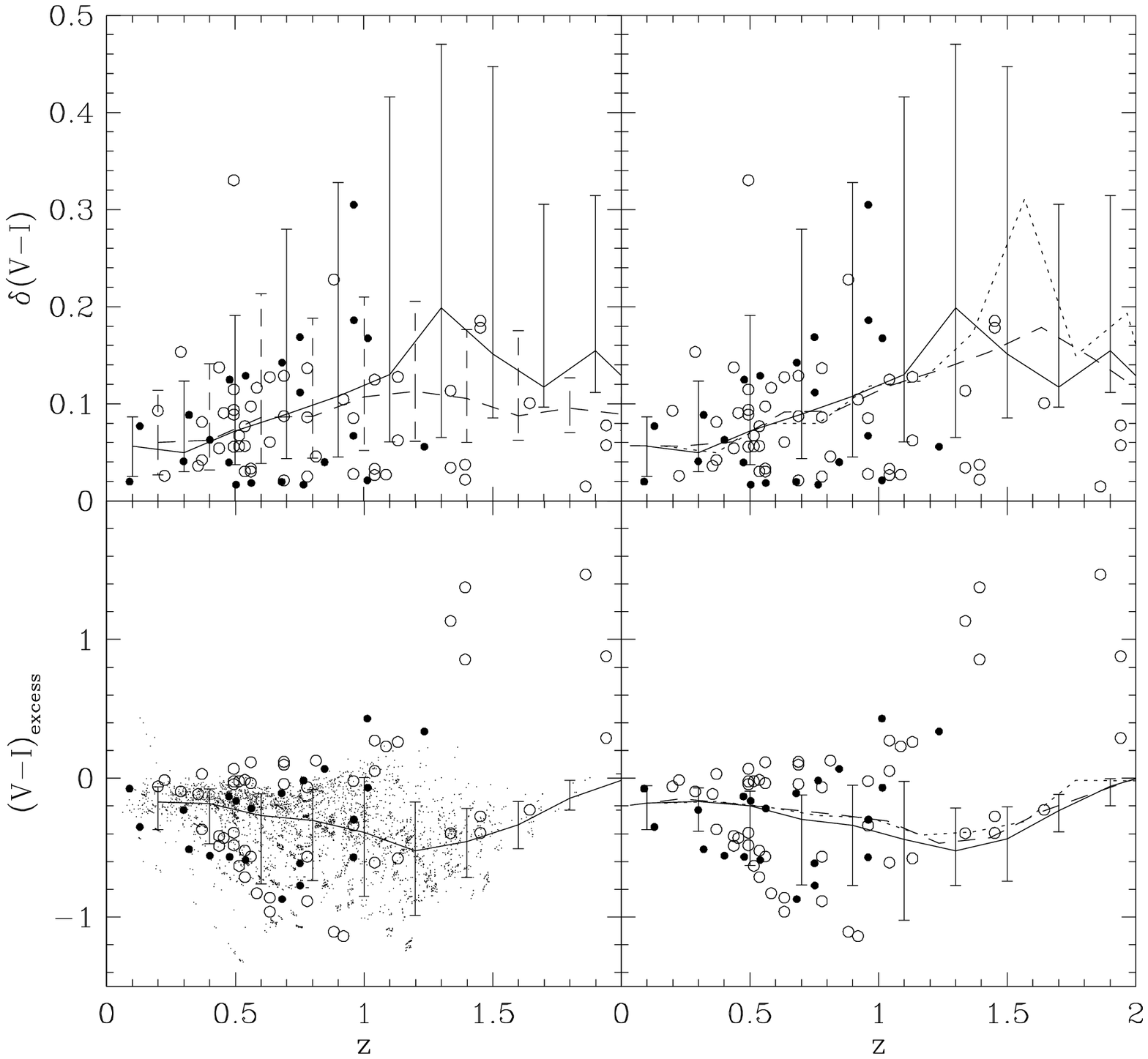,width=160mm,bbllx=0mm,bblly=95mm,bburx=190mm,bbury=270mm,clip=}
\caption{The variation of our two statistics with redshift. Top and
bottom rows show \index and \excess\ respectively. In each case
circles show HDF galaxies (filled points are spectroscopic redshifts,
open points are photometric redshifts). Lines show the median of the
model distribution with error bars indicating the 10\% and 90\%
intervals. In the left-hand column the solid line shows results for
the Hernquist profile (the dots show a sample of model galaxies to
give a more detailed indication of the distribution of model results),
while the dashed line shows the result for the $r^{1/4}$-law
profile. In the right-hand column the solid lines show results for our
default parameters, while dashed and dotted show results for the
alternative B/T$_{\rm S}$ and alternative $N_{\rm relax}$
respectively.}
\label{fig:statsvsz}
\end{figure*}

Finally, in Fig.~\ref{fig:statsvsz}, we compare model and data as a
function of redshift. For the model the median relation is shown by
the line, with error bars showing the 10\% and 90\% intervals at each
redshift. Error bars are typically shown on only one line per panel to
aid clarity but are of comparable size for the other lines. Data are
shown by circles; filled and open to indicate spectroscopic and
photometric redshifts respectively. As shown in the upper panels, both
model and data show an increasing median \index\ with redshift out to
$z\approx 1$. The top left-hand panel clearly demonstrates that the
model predictions for \index\ at high redshifts depend crucially on
the surface brightness profile adopted for the post-merger spheroidal,
with the Hernquist profile producing many more high \index\ galaxies
at high redshift than the $r^{1/4}$ profile (for which the
distribution of \index\ is in rather good agreement with the data at
all redshifts) The exact choice of model parameters has little effect
on the prediction (although increasing $N_{\rm relax}$ to 30 does
reduce the 90\% intervals).

In \excess\ (lower panels) the model spans a similar range to the data
as a function of redshift, the rapid increase in blue excess objects
between $z=0$ and $z=0.5$ is apparent in both model and data for
example. Interestingly, the data appears almost bimodal at these
redshifts, with one group of galaxies lying along $\excess\approx 0$
(and so are presumably old, passive spheroidals) and another group
lying along a locus of \excess\ which decreases with redshift. Dots
show a sample of model galaxies for comparison. Although a clear
branch of model points can be seen running along $\excess\approx 0$,
no well-defined branch is seen in the vicinity of the second group of
data points, although the model points certainly span the range of
blue \excess\ values seen in the data. Given the small number of data
points (and the fact that most of those in the second branch have only
photometric redshifts) no strong conclusions can be drawn as yet. Of
course, the model does not contain the very red galaxies seen in the
data, as expected from Fig.~\ref{fig:passive}.

Since we can easily switch off the effects of the PSF on model galaxies 
we have checked that the wavelength dependence of the PSF does
not significantly affect the distribution of \index\ for model
galaxies. Without the PSF the distribution of \index\ for model
galaxies is significantly higher in the lowest bin of \index\ shown in
Fig.~\ref{fig:statistics} (the wavelength dependence of the PSF
increases \index\ for a few of these galaxies, mostly at low
redshifts, giving them $\index \approx 0.1$) and the tail extends to
slightly higher \index\ (here it is the homogenising effects of the
PSF which are important, particularly for high redshift galaxies).

We have also checked the effect of the disk component of the model
spheroidal galaxies. Recall that our E/S0 class contains galaxies with
at least 44\% of their light in a spheroidal component, allowing up to
56\% to be in a disk (although many have much less than this of
course). The disk light can contribute to \index\ and \excess\ since
it may be of a different colour to the spheroidal light and is also
distributed differently. The broad features (e.g. width) of the
\index\ and \excess\ distributions remain the same whether we include
the disk components of E/S0 galaxies or not.

Without the disk component we find that the tail of the \index\
distribution extends to slightly higher values (in galaxies with a
recent burst adding a disk component can actually reduce \index\ since
the disk may be of similar colour to the burst component), but the
effect is small relative to the current theoretical and observational
uncertainties. In \excess\ leaving out the disk components reduces the
number of galaxies at $\excess \approx -0.4$ to $-0.1$, since model
disks often have \excess\ in this range. In conclusion, including the
disk component of our spheroidal galaxies has some effect on the
predicted distributions, but, with the present dataset, our basic
conclusions are insensitive to the presence or absence of a disk
component in model galaxies.

So far we have considered most of the parameters of the model to be
fixed at the values chosen by \scite{cole00}, the exceptions being
B/T$_{\rm S}$ and $N_{\rm relax}$ which we have constrained from
morphologically selected number counts and redshift distributions. By
adopting this philosophy the model results are true predictions, since
the model parameters were fixed by considerations entirely unrelated
to star formation activity in spheroidal galaxies.

Finally, we now consider the effect of altering key model parameters
which control the formation and merging histories of model galaxies.
We will not explore this graphically but it is interesting to examine
the consequences qualitatively for two reasons. Firstly, this will aid
our understanding of what physical processes are crucial for
constraining the statistics we have considered. Secondly, and most
importantly, it will test the ability of the type of comparison between
theory and observation developed in this paper to place useful
constraints on the formation rates of spheroidal galaxies. An important
caveat must be borne in mind --- namely that after we alter these
parameters it is unlikely that the model will be in as good agreement
with the morphologically selected counts and redshift distributions of
Figs.~\ref{fig:counts} \& \ref{fig:dndz} as with the standard set of
parameters. Such models are not necessarily physically realistic, and
we explore them only to demonstrate the sensitivity of \index\ and
\excess\ to the formation histories of spheroidals.

Three model parameters are crucial to the formation of spheroidals in
the model of \scite{cole00}. The frequency of mergers is controlled by
the parameter $f_{\rm df}$ which determines the strength of the
dynamical friction force felt by galaxies. The division between minor
and major mergers is determined by the parameter $f_{\rm ellip}$ and
the timescale of the star burst associated with a major merger depends
upon $\epsilon_{\star,{\rm burst}}$. (See \S\ref{sec:define} and
\scite{cole00} for further details.) Increasing $f_{\rm df}$ enlarges
the merger timescale and so reduces the frequency of the major mergers.
Increasing this parameter from 1 to 2 or 4 substantially reduces the
number of spheroidal galaxies. The fractional reduction is greatest for
high \index\ (highly negative \excess) galaxies. This arises since the
reduced rate of major mergers means that a given spheroid tends to have
experienced its last burst of star formation longer ago, which results
in the characteristic colours of the burst stars fading away before the
galaxy is observed.

Increasing $f_{\rm ellip}$ to 0.5 or 0.75 also has a large effect on
the distributions of \index\ and \excess. Firstly, there is an overall
reduction in the number of spheroidals, since major mergers become much
rarer. There is also some differential effect, with high
\index\ (equivalently, highly negative \excess) galaxies being depleted
preferentially, again because of the greater time since the most recent
merger. A secondary effect of the reduced merger rate is that,
particularly at low redshift, galaxies in major mergers tend to be less
gas rich when $f_{\rm ellip}$ is increased, as they have had much more
time to turn gas into stars quiescently. This also tends to reduce the
number of high \index\ (highly negative \excess) galaxies.

The parameter $\epsilon_{\star,{\rm burst}}$ has little effect on the
distributions of \index\ and \excess\ since it does not change the
frequency of bursts nor the mass of stars formed within them. It does
however alter the sensitivity of the results to the value of $N_{\rm
relax}$. For smaller $\epsilon_{\star,{\rm burst}}$ we find that
increasing $N_{\rm relax}$ produces a greater reduction in the numbers
of high \index\ and highly negative \excess\ galaxies. This occurs
because, with the slower star formation rates implied by a low
$\epsilon_{\star,{\rm burst}}$, bursts tend to redden more rapidly and
so become indistinguishable from the rest of their galaxy. With
$N_{\rm relax}=3$ many model spheroidals still have recognisable blue
colours to their cores, but for $N_{\rm relax}=30$ most have already
reddened sufficiently to deplete the highly negative regions of the
\excess\ distribution.

The \scite{cole00} model assumes that in the case of a minor merger
($M_1/M_2 < f_{\rm ellip}$) stars from the smaller galaxy are added to
the bulge of the larger galaxy, thereby providing a mechanism by which
bulges can form without any associated burst of star formation. An
alternative would be to assume the stars are added to the disk of the
larger galaxy, in which case spheroids can only ever form through a
major merger. \scite{cole00} found that the choice of where to place
stars from minor mergers made little difference to the properties of
model galaxies which they considered. We find a similar result, with
no significant difference being made to the \index\ and \excess\
distributions or their redshift dependence.

\section{Discussion}
\label{sec:discuss}

We have assessed the implications of the sample of elliptical galaxies
in the HDFs with inhomogeneous internal colours \cite{felipe} for the
picture of continuous formation of ellipticals expected in
hierarchical models of structure formation.

We have shown that the galaxy formation model of \scite{cole00} can
produce good agreement with data on morphologically selected number
counts, and produces many of the trends observed in morphologically
selected redshift distributions, although the detailed agreement in
these quantities is less good. In defining the morphology of model
galaxies we considered the time taken for a merger event to relax to
an elliptical galaxy configuration, quantifying this criterion in
terms of a multiple, $N_{\rm relax}$, of the elliptical's dynamical
time. Of crucial importance to the present work is that this
definition allows a galaxy to be classed as a relaxed elliptical
before the distinctive blue colour of recent star formation has
faded. With the present level of agreement between theory and
observations it is difficult to set strong constraints on $N_{\rm
relax}$, but $N_{\rm relax}\lsim 10$ is favoured, as would be expected
on theoretical grounds \cite{barnes92}. Accurate determinations of the
redshift distribution of elliptical galaxies will provide the
strongest constraint on this parameter.

We compare the observational data with model predictions through two
statistics, the inhomogeneity index originally defined by
\scite{felipe} and the difference between the observed (V-I) colour
and that of a passively evolving old stellar population at the same
redshift (which is simpler to compute from the model, but makes no use
of the spatially resolved information in the HDF images and which is
sensitive to errors in the photometric redshifts of HDF galaxies).

The theoretical predictions suffer from some uncertainties, due to
the rate at which ellipticals are expected to form in a given
cosmological model (as demonstrated in Fig.~\ref{fig:KCcompare}), the
uncertain spatial distribution of light from recent star formation for
which we have only limited theoretical guidance as yet and the
parameters used to specify what constitutes a spheroidal galaxy in the
model. Nevertheless, we have shown that our basic conclusions are
insensitive to the latter two uncertainties (the first is exactly what
we are attempting to measure from the data so necessarily has a strong
impact on the model predictions), allowing us to make a useful
comparison to the observational data.

Comparing our two statistics for model and observed galaxy samples we
find reasonable agreement in the distributions of both \index\ and \excess\
with two notable exceptions. Firstly, the current model produces too
few spheroidals with low \index\ and, secondly, the model does not
produce the population of $\excess > 0$ galaxies seen in the data. We
discuss these two points further below. However, most importantly for
the present work is the fact that the model does produce the correct
number of high \index\ and blue \excess\ galaxies. The implication of
this result is that a model in which elliptical galaxies form
hierarchically in a $\Lambda$CDM universe, and which was specifically
constrained to match a set of local data \cite{cole00}, is
quantitatively consistent with the continued star formation implied by
the HDF data of \scite{felipe} with no further adjustment of the free
parameters of the model. The basic trends of \index\ and \excess\ with
redshift are also reproduced by the model. An expanded data set will
allow these trends to be tested quantitatively.

However, there is evidence that the model produces
somewhat too few relaxed, homogeneous spheroidals from the \index\
statistic. Given the small sample size and the theoretical
uncertainties it is premature to assign much significance to this
apparent discrepancy. Perhaps the most noticeable discrepancy between
theory and observations is the lack of extremely red objects (relative
to a passively evolving spheroidal formed at high redshift) in the
model sample. The majority of these objects have only photometric
redshifts and so it remains possible that their apparently very red
colours for their redshift are simply a consequence of an incorrect
redshift determination. Both of these points should be addressed in
more detail, both using larger observational samples, and by
improvements in the theoretical calculations.

We have briefly considered the sensitivity of the \index\ and \excess\
statistics to the formation rate of spheroidals, by varying parameters
of the model which directly influence this formation rate. We
typically find that lower formation rates tend to preferentially
deplete the high \index\ and highly negative \excess\ regions of the
distributions. This is due mainly to the fact that, for a lower
spheroidal formation rate, a galaxy's last major merger tends to have
occured longer ago and so the characteristic colours of the recent
burst have faded away by the time it is observed. Thefore, analysis of
large samples of spheroidals using these techniques should allow us to
directly constrain their formation rate.

Spectroscopic observations of these faint populations of ellipticals
should allow estimates of their star formation rates and the ages of
their stellar populations. These quantities, for which we have
presented predictions, can be compared to the theoretical predictions
much more directly, and so should provide much stronger constraints on
the formation of ellipticals. With the imminent launch of the
Advanced Camera for Surveys on Hubble Space Telescope, there is
a strong case for assembling a large sample of field spheroidals
with redshifts and reliably-sampled internal colors.

\section*{Acknowledgements}

We acknowledge valuable discussions with Bob Abraham, Pieter van
Dokkum and Tomasso Treu. AJB thanks his collaborators Carlton Baugh,
Shaun Cole Carlos Frenk and Cedric Lacey for free use of the {\sc
galform} galaxy formation code.


\begin{thebibliography}{}
\bibitem[Abraham et al. <1996>]{abraham96}Abraham~R.~G., van den Bergh~S., Glazebrook~K., Ellis~R.~S., Santiago~B.~X., Surma~P., Griffiths~R.~E., 1996, ApJS, 107, 1
\bibitem[Barger et al. <1999>]{barger99}Barger~A.~J., Cowie~L.~L., Trentham~N., Fulton~E., Hu~E.~M., Songaila~A., Hall~D., 1999, AJ, 117, 102
\bibitem[Barnes \& Hernquist <1992>]{barnes92}Barnes~J.~E., Hernquist~L., 1992, ARA\&A, 30, 705
\bibitem[Barnes \& Hernquist <1996>]{barnes96}Barnes~J.~E., Hernquist~L., 1996, ApJ, 471, 115
\bibitem[Barnes <1998>]{barnes98}Barnes~J.~E., 1998, in Galaxies:
Interactions and Induced Star Formation, Saas-Fee Advanced Course
26. Lecture Notes 1996. Swiss Society for Astrophysics and Astronomy,
XIV, Edited by R.~C.~Kennicutt, Jr., F.~Schweizer, J.~E.~Barnes,
D.~Friedli, L.~Martinet and D.~Pfenniger. Springer-Verlag
Berlin/Heidelberg, p. 275.
\bibitem[Baugh, Cole \& Frenk <1996a,b>]{baugh96dummy}
\bibitem[Baugh, Cole \& Frenk <1996a>]{baugh96a}Baugh~C.~M., Cole~S., Frenk~C.~S., 1996a, MNRAS, 282, L27
\bibitem[Baugh, Cole \& Frenk <1996b>]{baugh96b}Baugh~C.~M., Cole~S., Frenk~C.~S., 1996b, MNRAS, 283, 1361
\bibitem[Baugh et al. <1998>]{baugh98}Baugh~C.~M., Cole~S., Frenk~C.~S., Lacey~C.~G., 1998, ApJ, 498, 504
\bibitem[Bower et al. <1990>]{bower90}Bower~R.~G., Ellis~R.~S., Rose~J.~A., Sharples~R.~M., 1990, AJ, 99, 530
\bibitem[Brinchmann <1999>]{brinchmann00}Brinchmann~J., 1999, Ph.D. thesis, Univ. Cambridge
\bibitem[Casertano et al. <2000>]{casertano00}Casertano~S., de Mello~D., Dickinson~M., Ferguson~H.~C., Fruchter~A.~S., Gonzalez-Lopezlira~R.~A., Heyer~I., Hook~R.~N., Levay~Z., Lucas~R.~A., Mack~J., Makidon~R.~B., Mutchler~M., Smith~T., Stiavelli~M., Wiggs~M.~S., Williams~R.~E. ,2000, ApJ, 120, 2747
\bibitem[Cole et al. <2000>]{cole00}Cole~S., Lacey~C.~G., Baugh~C.~M., Frenk~C.~S., 2000, MNRAS in press
\bibitem[Daddi et al. <2000>]{daddi00}Daddi~E., Cimatti~A., Renzini~A., 2000, A\&A, 362, 45
\bibitem[Driver et al. <1995a, 1995b>]{driver95dummy}
\bibitem[Driver et al. <1995a>]{driver95a}Driver~S.~P., Windhorst~R.~A., Ostrander~E.~J., Keel~W.~C., Griffiths~R.~E., Ratnatunga~K.~U., 1995a, ApJ, 449, L23
\bibitem[Driver et al. <1995b>]{driver95b}Driver~S.~P., Windhorst~R.~A., Griffiths~R.~E., 1995b, ApJ, 453, 48
\bibitem[Efstathiou, Lake \& Negroponte <1982>]{eln82}Efstathiou~G., Lake~G., Negroponte~J., 1982, MNRAS, 199, 1069
\bibitem[Efstathiou \& Silk <1983>]{efstath83}Efstathiou~G., Silk~J., 1983, Fundamentals of Cosmic Physics, 9, 1
\bibitem[Eggen, Lynden-Bell \& Sandage <1962>]{eggen62}Eggen~O.~J., Lynden-Bell~D., Sandage~A.~R., 1962, ApJ, 136, 748
\bibitem[Ellis et al. <1997>]{ellis97}Ellis~R.~S., Smail~I., Dressler~A., Couch~W.~J., Oemler~A., Butcher~H., Sharples~R.~M., 1997, ApJ, 483, 582
\bibitem[Firth et al. <2001>]{firth01}Firth~A.~E. et al., 2001, astro-ph/0108182 (submitted to MNRAS)
\bibitem[Franceschini et al. <1998>]{frances99}Franceschini~A., Silva~L., Fasano~G., Granato~L., Bressan~A., Arnouts~S., Danese~L., 1998, ApJ, 506, 600
\bibitem[Glazebrook et al. <1995>]{glazebrook95}Glazebrook~K., Ellis~R., Santiago~B., Griffiths~R., 1995, MNRAS, 275, L19
\bibitem[Granato et al. <2000>]{granato00}Granato~G.~L., Lacey~C.~G., Silva~L., Bressan~A., Baugh~C.~M., Cole~S., Frenk~C.~S., 2000, ApJ, 542, 710
\bibitem[Hernquist <1992>]{hernquist92}Hernquist~L., 1992, ApJ, 400, 460
\bibitem[Im et al. <1999>]{im01a}Im~M., Griffiths~R.~E., Naim~A., Ratnatunga~K.~U., Roche~N., Green~R.~F., Sarajedini~V.~L., 1999, ApJ, 510, 821
\bibitem[Im et al. <2001>]{im01b}Im~M., Simard~L., Faber~S.~M., Koo~D.~C., Gebhardt~K., Willmer~C.~N.~A., Phillips~A., Illingworth~G., Vogt~N.~P.,Sarajedini~V.~L., 2001, astro-ph/0011092
\bibitem[Kauffmann, White \& Guiderdoni <1993>]{kauff93}Kauffmann~G., White~S.~D.~M., Guiderdoni~B., 1993, MNRAS, 264, 201
\bibitem[Kauffmann <1996>]{kauff96}Kauffmann~G., 1996, MNRAS, 281, 487
\bibitem[Kauffmann \& Charlot <1998a>]{kauff98}Kauffmann~G., Charlot~S., 1998a, MNRAS, 294, 705
\bibitem[Kauffmann et al. <1999>]{kauff99}Kauffmann~G., Colberg~J.~M., Diaferio~A., White~S.~D.~D., 1999, MNRAS, 303, 188
\bibitem[Lacey et al. <2001>]{laceyinprep}Lacey~C.~G. et al., 2001, in preparation
\bibitem[Marzke et al. <1998>]{marzke98}Marzke~R.~O., da Costa~L.~N., Pellegrini~P.~S., Willmer~C.~N.~A., Geller~M.~J., 1998, ApJ, 503, 617
\bibitem[McCarthy et al. <2001>]{mccarthy00}McCarthy~P.~J. et al., 2001, astro-ph/0108171 (submitted to ApJ)
\bibitem[Menanteau et al. <1999>]{menanteau99}Menanteau~F., Ellis~R.~S., Abraham~R.~G., Barger~A.~J., Cowie~L.~L., 1999, MNRAS, 309, 208
\bibitem[Menanteau, Abraham \& Ellis <2001>]{felipe}Menanteau~F., Abraham~R.~G., Ellis~R.~S., 2001, MNRAS, 322, 1
\bibitem[Menanteau, Jimenez \& Matteucci <2001>]{fmnraul}Menanteau~F., Jimenez~R., Matteucci~F., 2001, submitted to ApJL
\bibitem[Mo, Mao \& White <1998>]{mmw98}Mo~H.~J., Mao~S., White~S.~D.~M., 1998, MNRAS, 295, 319
\bibitem[Sandage \& Visvanathan <1978>]{sandage78}Sandage~A., Visvanathan~N., 1978, ApJ, 223, 707
\bibitem[Sanders \& Mirabel <1996>]{sanders96}Sanders~D.~B., Mirabel~I.~F., 1996, ARA\&A, 34, 749
\bibitem[Schade et al. <1999>]{schade99}Schade~D., Lilly~S.~J., Crampton~D., Ellis~R.~S., Le F\`evre~O., Hammer~F., Brinchmann~J., Abraham~R., Colless~M., Glazebrook~K., Tresse~L., Broadhurst~T., 1999, ApJ, 525, 31
\bibitem[Somerville, Primack \& Faber <2001>]{somerville01}Somerville~R.~S., Primack~J.~R., Faber~S.~M., 2001, MNRAS, 320, 504
\bibitem[Stanford, Eisenhardt \& Dickinson <1998>]{stanford98}Stanford~S.~A., Eisenhardt~P.~R., Dickinson~M., 1998, ApJ, 492, 461
\bibitem[Treu et al. <2001>]{treu01}Treu~T., Stiavelli~M., Bertin~G., Casertano~S., M\/oller~P., 2001, astro-ph/0106147
\bibitem[van Dokkum et al. <1998, 2000>]{vandokdummy}
\bibitem[van Dokkum et al. <1998>]{vandok98}van Dokkum~P.~G., Franx~M., Kelson~D.~D., Illingworth~G.~D., 1998, ApJ, 504, 17
\bibitem[van Dokkum et al. <2000>]{vandok00}van Dokkum~P.~G., Franx~M., Fabricant~D., Illingworth~G.~D., Kelson~D.~D., 2000, ApJ, 541, 95
\bibitem[van Dokkum et al. <2001>]{vandokkum01}van Dokkum~P.~G., Franx~M., Kelson~D.~D., Illingworth~G.~D. 2001, ApJ, 553, 39
\bibitem[Walker, Mihos \& Hernquist <1996>]{walker96}Walker~I., Mihos~J.~C., Hernquist~L., 1996, ApJ, 460, 121
\bibitem[Williams et al. <1996>]{williams96}Williams~R.~E. et al., 1996, AJ, 112, 1335
\bibitem[Zepf <1997>]{zepf97}Zepf~S.~E., 1997, Nat., 390, 377
\bibitem[Zucca et al. <1997>]{zucca97}Zucca~E. et al., 1997, A\&A, 326, 477
\end{thebibliography}
\end{document}